\documentclass{jfm}
\usepackage{graphicx}
\usepackage{epstopdf, epsfig}
\usepackage{natbib}
\usepackage{amssymb}
\usepackage{amsmath}
\usepackage{bm} 
\usepackage{latexsym}
\usepackage{lineno}

\newcommand\Reytau{\mathrm{Re}_{\tau}}
\newcommand\dd{\mathrm{d}}

\shorttitle{stream-wise Reynolds stress}

\shortauthor{P. A. Monkewitz}

\title{Asymptotics of stream-wise Reynolds stress in wall turbulence}

\author{Peter A. Monkewitz
  \corresp{\email{peter.monkewitz@epfl.ch}}}

\affiliation{\'Ecole Polytechnique F\'ed\'erale de Lausanne (EPFL), CH-1015, Lausanne, Switzerland}

\begin{document}

\maketitle   

\begin{abstract}
The scaling of different features of stream-wise normal stress profiles $\langle uu\rangle^+(y^+)$ in turbulent wall-bounded flows, in particular in truly parallel flows, such as channel and pipe flows, is the subject of a long running debate. Particular points of contention are the scaling of the ``inner'' and ``outer'' peaks of $\langle uu\rangle^+$ at $y^+\approxeq 15$ and $y^+ =\mathcal{O}(10^3)$, respectively, their infinite Reynolds number limit, and the rate of logarithmic decay in the outer part of the flow. Inspired by the landmark paper of \citet{chen_sreeni2021}, two terms of the inner asymptotic expansion of $\langle uu\rangle^+$ in the small parameter $\Reytau^{-1/4}$ are extracted for the first time from a set of direct numerical simulations (DNS) of channel flow. This inner expansion is completed by a matching outer expansion, which not only fits the same set of channel DNS within 1.5\% of the peak stress, but also provides a good match of laboratory data in pipes and the near-wall part of boundary layers, up to the highest $\Reytau$'s of order $10^5$. The salient features of the new composite expansion are first, an inner $\langle uu\rangle^+$ peak, which saturates at 11.3 and decreases as $\Reytau^{-1/4}$, followed by a short ``wall loglaw'' with a slope that becomes positive for $\Reytau \gtrapprox 20'000$, leading up to an outer peak, and an outer logarithmic overlap with a negative slope continuously going to zero for $\Reytau\to\infty$.
\end{abstract}

\section{\label{sec1}Introduction}

In the following, the classical non-dimensionalization is adopted with the ``inner'' or viscous length scale $\widehat{\ell} \equiv (\widehat{\nu}/\widehat{u}_\tau)$, and $\widehat{u}_\tau \equiv (\widehat{\tau}_w/\widehat{\rho})^{1/2}$, $\widehat{\rho}$ and $\widehat{\nu}$ are the friction velocity, density and dynamic viscosity, respectively, with hats identifying dimensional quantities. The resulting non-dimensional inner and outer wall-normal coordinates are $y^+=\widehat{y}/\widehat{\ell}$ and $Y=y^+/\Reytau$, respectively, with $\Reytau\equiv \widehat{L}/\widehat{\ell}$ the friction Reynolds number.

The scaling of normal Reynolds stresses in turbulent boundary layers, in particular of the stream-wise component $\langle uu\rangle^+$, which is experimentally accessible with single hotwires, has been the subject of a long debate. A mixed scaling with $\widehat{u}_\tau \widehat{U}_\infty$, proposed by \citet{DE2000}, did somewhat improve the collapse of normal stress profiles in turbulent flat-plate boundary layers at different Reynolds numbers, but the result was still far from satisfactory, as seen for instance in \citet{Chauhan09}. Without much theoretical support available, some groups have turned to models and in particular to the attached eddy model, originally proposed by \citet{Towns56}. This model was subsequently developed in Melbourne by \citet{PHC86} and \citet{PM95}, among others, and, since then, refined by the ``Melbourne school'', well represented in the author list of \citet{samie_etal_2018}. Here, only the salient features of the model, recently reviewed by \citet{MarusicMonty19}, are summarized: its linchpin is the unlimited increase with Reynolds number of the peak of the inner-scaled $\langle uu\rangle^+$ around $y^+\approxeq 15$, given as $3.54 + 0.646 \ln\Reytau$ in \citet{samie_etal_2018}. If true for arbitrarily large $\Reytau$, this unlimited growth of the inner peak means that the inner scaling of $\langle \widehat{u}\widehat{u}\rangle$ is not appropriate and must be replaced by mixed scaling \citep[see also][]{TSFP10Monk}.
Closely related to the unlimited growth of the inner peak of $\langle uu\rangle^+$, predicted by the attached eddy model, is the universal outer logarithmic decrease of $\langle uu\rangle^+$ towards the centerline, as $1.95 - 1.26\,\ln Y$ \citep[see e.g.][]{MMHS13}. As pointed out by \citet{MonkNagib2015}, the constancy of the outer decay rate hinges on the unlimited growth of $\langle uu\rangle^+$ near the wall. For the zero pressure gradient flat plate turbulent boundary layer, henceforth abbreviated ZPG TBL, \citet{MonkNagib2015} have shown, that an unlimited growth of the inner $\langle uu\rangle^+$ peak is incompatible with the Taylor expansion of the full stream-wise mean momentum equation about $y^+=0$ \citep[see also the discussion by][]{chen_sreeni2021}.

At this point it should be clear, that the difficulty of identifying the asymptotic scalings of the different parts of $\langle uu\rangle^+$ stems from the variation of its main features with $\Reytau^{1/4}$ or $\ln\Reytau$ : While the experimental difficulty of measuring $\langle uu\rangle^+$ at $\Reytau = 10^5$ is indeed ``extreme'' \citep{Hultetal12}, both $\Reytau^{1/4}$ and $\ln\Reytau$ increase by a factor of less than 5 between $\Reytau = 10^3$ and $10^5$. This calls for the machinery of matched asymptotic expansions \citep[see e.g.][]{KC85}, henceforth abbreviated MAE, which is even more important than for the analysis of mean velocity profiles, because modest changes of $\Reytau^{1/4}$ or $\ln \Reytau$ entrain substantial variations of $\langle uu\rangle^+$.

The program of the paper is to first extract the inner asymptotic expansion of the stream-wise normal stress from DNS of channel flow in section \ref{sec2} and to complete the composite expansion of $\langle uu\rangle^+$ with the outer expansion in section \ref{sec3}. Conclusions and comments are collected in section \ref{sec4}.

\section{\label{sec2}The inner expansion of the streamwise normal stress $\langle uu\rangle^+$}

The inner asymptotic expansion of $\langle uu\rangle^+$ for large $\Reytau$ is extracted from the channel DNS of table \ref{TableDNS} in a similar fashion as the mean velocity expansion in \citet{Monk21}.

\begin{table}
\center
\caption{Channel DNS profiles used to determine contributions of $\mathcal{O}(\ln^n\Reytau)$}
\begin{tabular}{l l l l}
\hline
Profile & $\Reytau$ & Color in figs. & Reference \\
\hline
\#1 & 5186 & \quad red & \citet{LM15} \\
\#2 & 3000 & \quad pink & \citet{TMG13} \\
\#3 & 2004 & \quad violet & \citet{HJ06} \\
\#4 & 1995 & \quad blue & \citet{LM15} \\
\#5 & 1000 & \quad green & \citet{LM15} \\
\label{TableDNS}
\end{tabular}
\end{table}

As mentioned in the introduction, the inner, near-wall peak $\langle uu\rangle^+_\mathrm{IP}$ at $y^+\approxeq 15$ undergoes a significant growth with $\Reytau$, but \citet{chen_sreeni2021} argued, based on the exact maximum of 1/4 for the turbulent energy production, that the inner peak height remains finite and decreases as $\Reytau^{-1/4}$. They also showed in their appendix that this $\Reytau^{-1/4}$ scaling is robust, because the terms of the Taylor expansion of $\langle uu\rangle^+$ beyond $(y^+)^2$, that were neglected in the derivation, can be accounted for by a proportionality factor \citep[equ. 5.9a,b of][]{chen_sreeni2021}. The upper limit of 1/4 for the coefficient of $(y^+)^2$ in the Taylor expansion of $\langle uu\rangle^+$ about the wall is consistent, within uncertainty, with the value of 0.26, extrapolated by \citet[fig. 6 and equ. 2.19]{MonkNagib2015} from ZPG TBL DNS.

The findings of \citet{chen_sreeni2021} are fully confirmed by the present analysis of the DNS data in table \ref{TableDNS}. Using the technique described in section 3.1 of \citet{Monk21}, the decomposition
\begin{equation}
\label{decomp}
\langle uu\rangle^+_{\mathrm{DNS}} = f(y^+) + \Phi(\Reytau) g(y^+)
\end{equation}
is obtained from different pairs of DNS profiles, and for different gauge functions $\Phi$. It turns out, that the only $\Phi$'s which produce a satisfactory collapse of the $f$'s and $g$'s obtained from all possible profile pairs in table \ref{TableDNS} are $\Phi = \Reytau^{-1/4}$ and $\Phi = 1/\ln \Reytau$. $\Phi = \ln \Reytau$, on the other hand, which corresponds to the inner peak scaling of the attached eddy model, produces \textbf{no sign of a collapse} of the $f$'s and $g$'s from different DNS pairs.

\begin{figure}
\center
\includegraphics[width=0.7\textwidth]{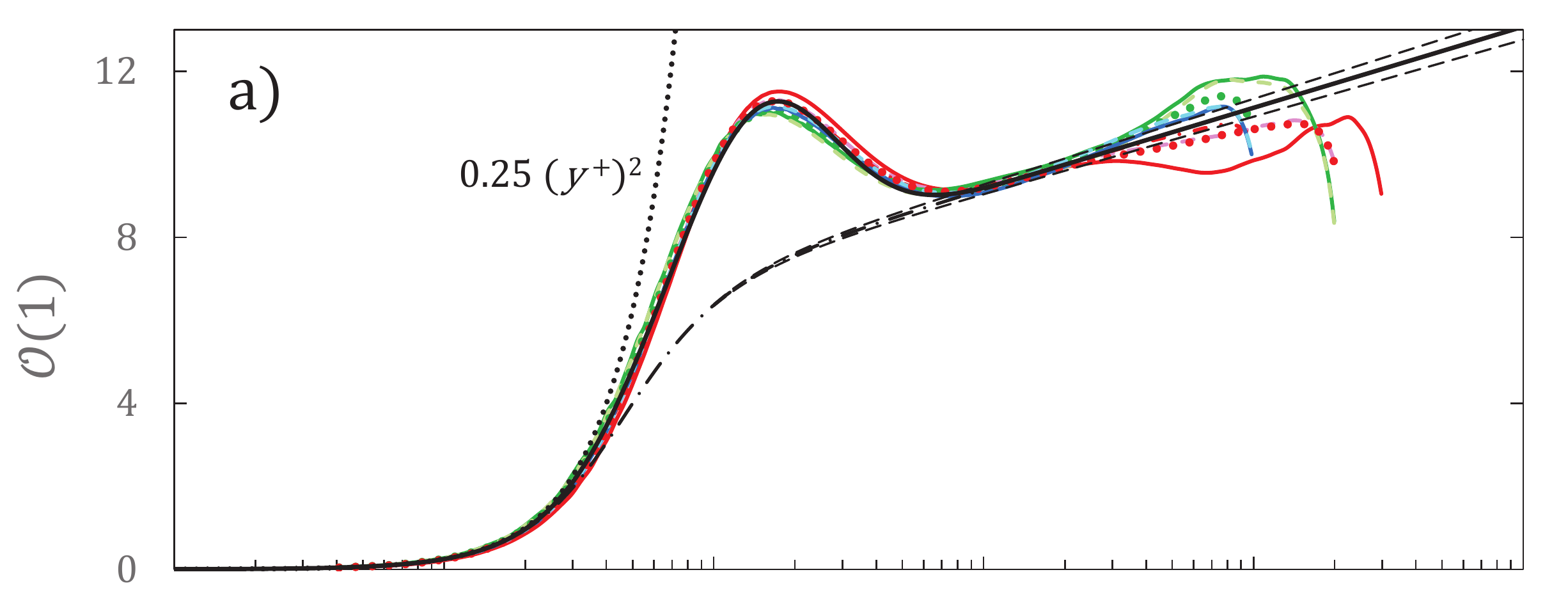}
\includegraphics[width=0.7\textwidth]{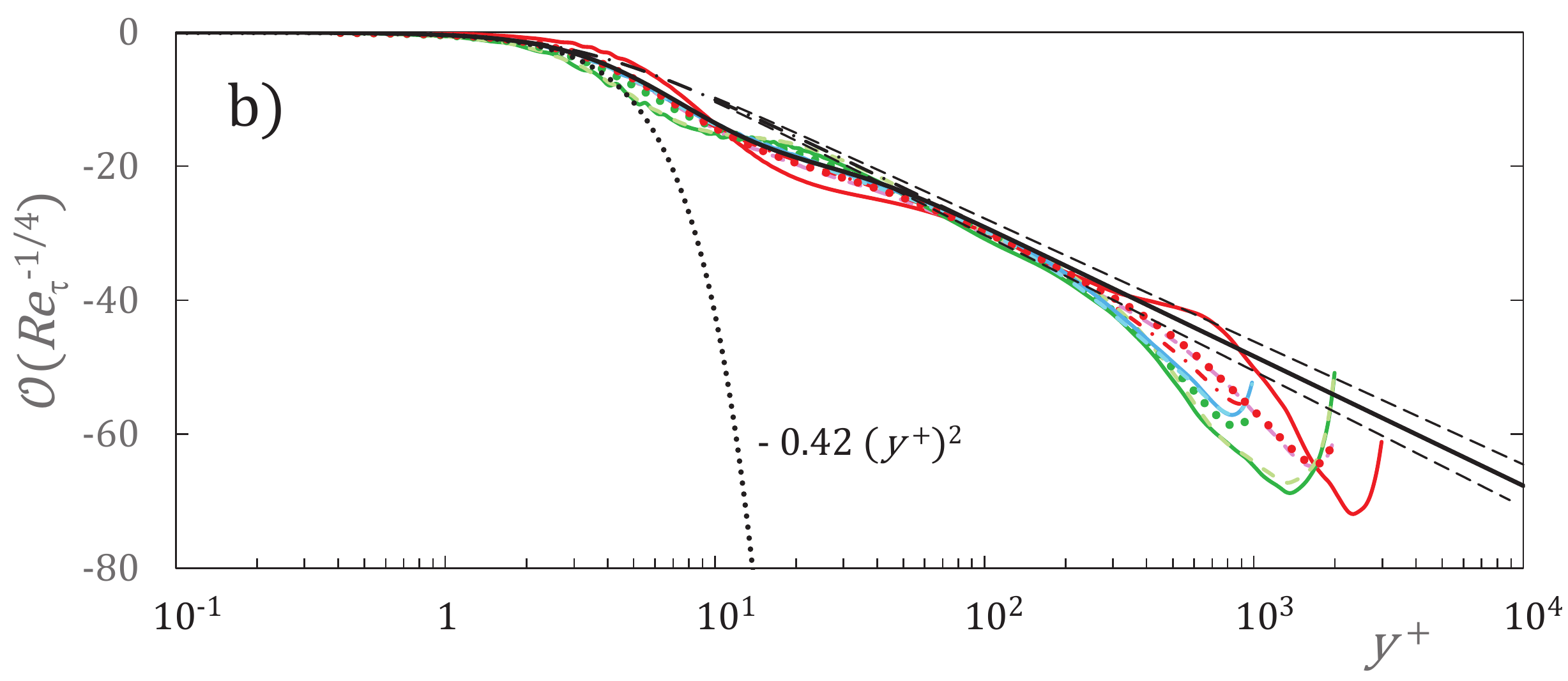}
\caption{\label{fig1} The $\mathcal{O}(1)$ (panel a) and $\mathcal{O}(1/\Reytau^{1/4})$ (panel b) components of $\langle uu\rangle^+$ extracted from DNS pairs of table \ref{TableDNS}: (red) profiles \#1\&2 (---), \#1\&3 (- - -),\#1\&4 ($\cdots$),\#1\&5 ($-\cdot-$); (green) \#2\&3 (---), \#2\&4 (- - -),\#2\&5 ($\cdots$); (blue) \#3\&5 (---), \#4\&5 (- - -). Panel (a): (black) --- and $-\cdot-$, $\mathcal{O}(1)$ part of equ. (\ref{uuwall}) with and without hump. Panel (b): (black) --- and $-\cdot-$, $\mathcal{O}(1/\Reytau^{1/4})$ part with and without hump. (black) - - -, logarithmic slopes modified by $\pm 5\%$.}
\end{figure}

In view of the theoretical underpinning of the $\Reytau^{-1/4}$ scaling by \citet{chen_sreeni2021}, only the results for the scaling $\Phi = \Reytau^{-1/4}$ are presented. The decomposition of $\langle uu\rangle^+_{\mathrm{DNS}}$ into a $\mathcal{O}(1)$ and a $\mathcal{O}(\Reytau^{-1/4})$ term is shown in figure \ref{fig1} and the collapse from different profile pairs is seen to be rather good up to $y^+$ of several hundred. Apart from the expected ``hump'' at $y^+\approxeq 15$, the decomposition (\ref{decomp}) reveals, for the first time, a short, but clear logarithmic region between $y^+\approx 60$ and 300, with a logarithmic slope that decreases as $\Reytau^{-1/4}$ from its maximum of 0.85 at infinite $\Reytau$ (fig \ref{fig1}a). This near-wall logarithmic region will prove to be essential for the construction of the complete inner asymptotic expansion.

In order to generate $\langle uu\rangle^+$ profiles for any Reynolds number, the fit $\mathcal{M}_2$ (equ. \ref{M2}), inspired by the construction of the Musker profile \citep{Musker79} for the mean profile, has been developed for the $\mathcal{O}(1)$ part $f$ and is supplemented by the ``hump'' function $\mathcal{H}$ (equ. \ref{Hump}). The $\mathcal{O}(\Reytau^{-1/4})$ part $g$ is fitted by the ``corner function'' $\mathcal{C}$ (equ. \ref{corner}). Hence, the near-wall stress, up to higher order terms (H.O.T.), is described by:
\begin{align}
\label{uuwall}
 \langle uu\rangle^+_{\mathrm{wall}} &= \mathcal{M}_2\left(y^+; 0.25, 373, 1.7, 8.8616\right) + \mathcal{H}(y^+; 4, 1.3, 15) + \nonumber \\
&+ \frac{1}{\Reytau^{1/4}}\,\Big\{\mathcal{C}(y^+; 3.1623, -8.4, 2) + \mathcal{H}(y^+; -3.7, 1, 13)\Big\} + \mathrm{H.O.T.}
\end{align}
To go beyond $\mathcal{O}(\Reytau^{-1/4})$, the Reynolds number range of available DNS and their mutual consistency are insufficient.

From equations (\ref{uuwall}) and (\ref{M2Taylor}),  one readily obtains the Taylor expansion of $\langle uu\rangle^+$ about the wall as
\begin{equation}
\label{uuTaylor}
\langle uu\rangle^+_{\mathrm{wall}}\,(y^+\to 0) = \left(0.25 - 0.42\, \Reytau^{-1/4}\right)(y^+)^2\,-\, 0.02\,\left(1 - \Reytau^{-1/4}\right)(y^+)^4 + ...
\end{equation}
It is noted, that the coefficient of $(y^+)^2$ in equation (\ref{uuTaylor}) fits the $\overline{b_1^2}$ in table 1 of \citet{hultlex2021} within less than $1\%$, which is not surprising as they used the same DNS data.

\begin{figure}
\center
\includegraphics[width=0.6\textwidth]{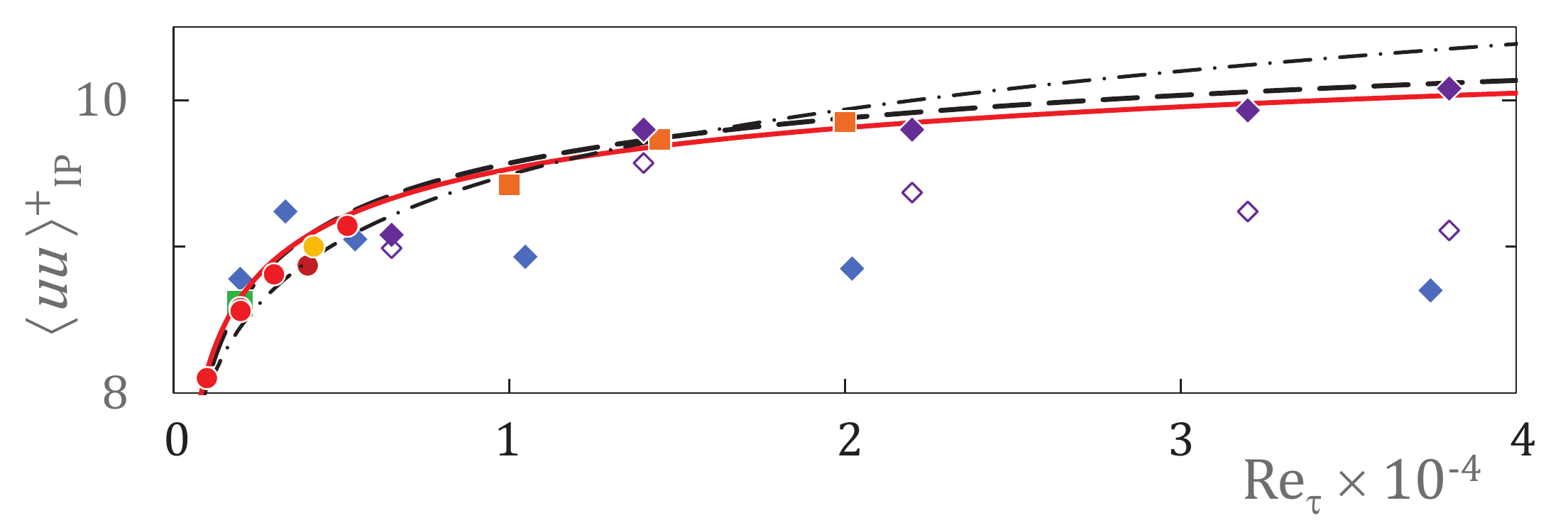}
\caption{\label{fig2} Inner peak height $\langle uu\rangle^+_{\mathrm{IP}}$ vs. $\Reytau$. (red) ---, equ. \ref{uuIP}; (black) - - -, $11.5 - 19.3\,\Reytau^{-1/4}$ of \citet{chen_sreeni2021}; (black) $-\cdot -$, $3.54 + 0.646\,\ln\Reytau$ of \citet{samie_etal_2018}. $\bullet$, channel DNS : (red) table \ref{TableDNS}, (dark red) \citet{bernardini_pirozzoli_orlandi_2014}, (yellow) \citet{LJ14}. $\blacklozenge$, pipe : (blue) Superpipe data of \citet{Hultetal12}, (purple) CICLoPE hotwire data of \citet{FioriniPhD} corrected according to \citet{smitsetal2011} ($\lozenge$, uncorrected). $\blacksquare$, ZPG TBL: (green) \citet{sillero2013}, (orange) \citet{samie_etal_2018}.}
\end{figure}

From equation (\ref{uuwall}) or figure \ref{fig1}, the inner peak height $\langle uu\rangle^+_{\mathrm{IP}}$ is obtained as
\begin{equation}
\label{uuIP}
\langle uu\rangle^+_{\mathrm{IP}}\,(y^+ \approxeq 15) = 11.3 - 17.7\, \Reytau^{-1/4} \quad ,
\end{equation}
to be compared in figure \ref{fig2} with $11.5 - 19.3\,\Reytau^{-1/4}$ of \citet{chen_sreeni2021}, with the correlation $3.54 + 0.646\,\ln\Reytau$ given by \citet{samie_etal_2018}, and with some data. Note that, while the latter correlation fits the data in figure \ref{fig2} quite well, the unlimited growth of $\langle uu\rangle^+_{\mathrm{IP}}$ with $\ln \Reytau$ is inconsistent with both the theoretical arguments of \citet{chen_sreeni2021} and the present decomposition (\ref{decomp}, \ref{uuwall}).

\begin{figure}
\center
\includegraphics[width=1.0\textwidth]{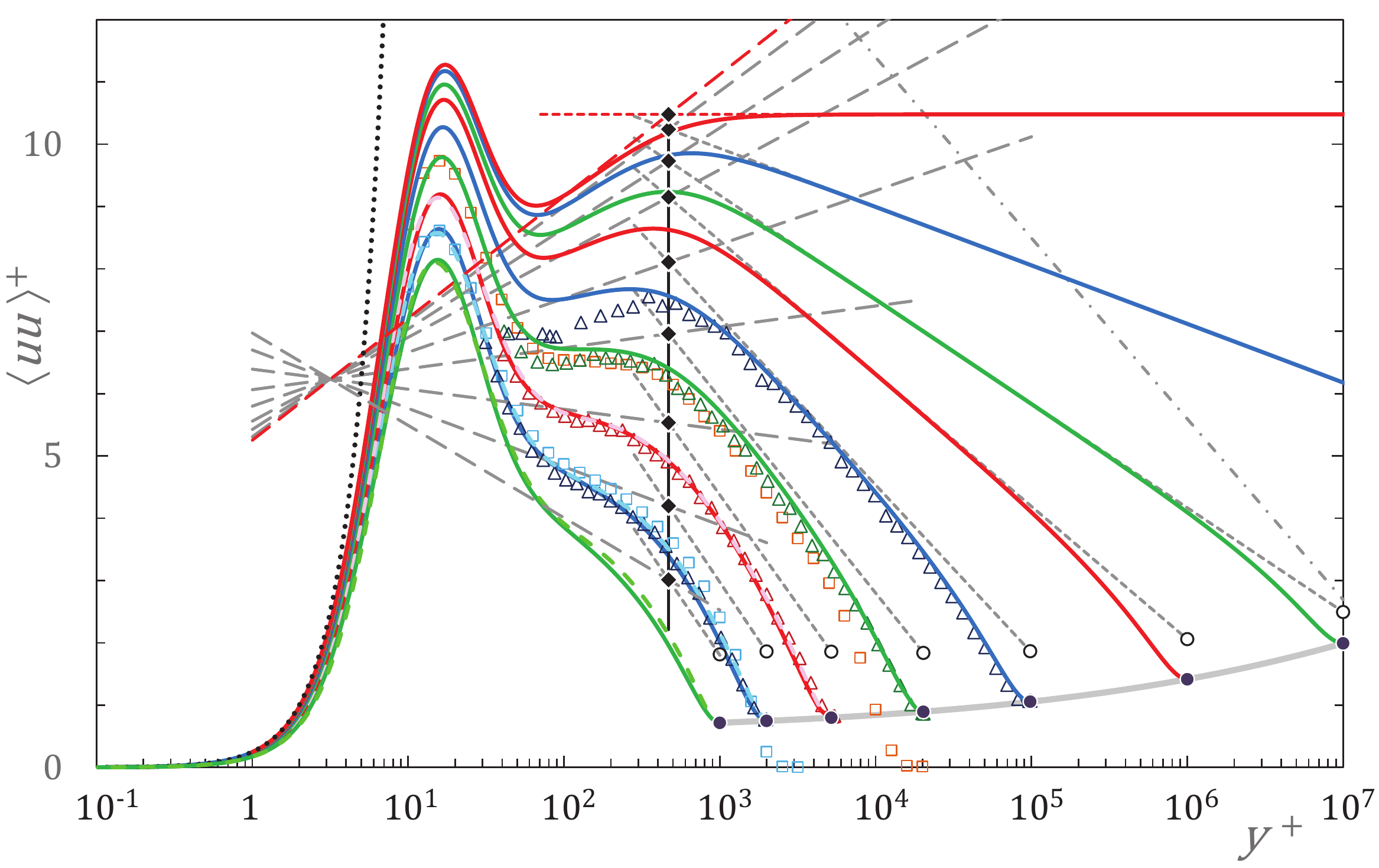}
\caption{\label{fig3} Solid lines: composite profiles of $\langle uu\rangle^+$ for $\Reytau$ = 1000 (green), 1995 (blue), 5186 (red), 20250 (green), 98190 (blue), $10^6$ (red), $10^7$ (green), $10^9$ (blue), $\infty$ (red). - - -, DNS profiles \#5 (green), \#4 (blue) and \#1 (red) of table \ref{TableDNS}. $\vartriangle\vartriangle\vartriangle$ , NSTAP Superpipe profiles of \citet{Hultetal12} at $\Reytau$ = 1985, 5411, 20250 and 98187. $\square$ (blue) , ZPG TBL profile of \citet{sillero2013}; $\square$ (orange) , ZPG TBL profile of \citet{samie_etal_2018} for $\Reytau$ = 14500. Long grey dashes (red for $\Reytau = \infty$), wall log laws (equ. \ref{uuwallinf}) for the 11 $\Reytau$ ; short grey dashes (red for $\Reytau = \infty$), corresponding overlap log laws (equ. \ref{uuwallinf}) with $\circ$ marking their ``end points'' at $y^+ = \Reytau$ ; $\blacklozenge$ , intersections of wall and overlap log laws at $y^+_\times$ (equ. \ref{yx}) . --- (gray), fit (\ref{uuCL}) of $\langle uu\rangle^+_{\mathrm{CL}}$, with $\bullet$ marking the fit at the $\Reytau$'s of the profiles shown. $\cdots$ , leading term $0.25\,(y^+)^2$ of the Taylor expansion about the wall. $- \cdot -$ , logarithmic slope of -1.26.}
\end{figure}

With equations (\ref{M2log}) and (\ref{corner}) of appendix \ref{App}, the large $y^+$ asymptote of the profile (\ref{uuwall}) is the logarithmic law
\begin{align}
\label{uuwallinf}
\langle uu\rangle^+_{\mathrm{wall}}(y^+\gg 1) =& \,S_{\mathrm{wall}}\,\ln y^+ \,+\,C_{\mathrm{wall}} \quad \mathrm{with} \\
S_{\mathrm{wall}} =& \,0.85 - 8.4\,\Reytau^{-1/4}\,,\quad C_{\mathrm{wall}} = 5.2513 + 9.6709\,\Reytau^{-1/4}
\label{uuwallcof}
\end{align}
The logarithmic asymptotes of $\langle uu\rangle^+_{\mathrm{wall}}$ for different $\Reytau$ are visualized in figure \ref{fig3} by the fan of straight lines intersecting at $y^+\approx 3$, and their logarithmic slope $S_{\mathrm{wall}}$, negative at low $\Reytau$, is seen to become positive at $\Reytau \approx 10^4$. This suggests a relation to the appearance of an outer peak in the $\langle uu\rangle^+$ profile, as discussed for instance by \citet{samie_etal_2018}. To actually form such an outer peak, the wall asymptote (\ref{uuwallinf}) has to cross over to a decay law at some $y^+_\times$.

This cross-over location $y^+_\times$ and the decay law beyond $y^+_\times$ could in principle be extracted from DNS data in a manner similar to the determination of $\langle uu\rangle^+_{\mathrm{wall}}$. Due to the complexity of the expansion and the limitations of the DNS data, this has not been possible. Still, figure \ref{fig1} shows that the channel DNS closely follows the wall loglaw (\ref{uuwallinf}) up to $y^+\approx 300$. Hence, $y^+_\times$ must be larger than 300, confirming the connection to the experimentally observed outer peak locations.

Turning to a straight fit of $y^+_\times$, all the data between $\Reytau = 10^3$ and $10^5$ are seen in figure \ref{fig3} to be compatible with $y^+_\times$ equal to a simple constant
\begin{equation}
\label{yx}
\ln y^+_\times = 6.15 \, , \, \langle uu\rangle^+_\times \equiv \langle uu\rangle^+_{\mathrm{wall}}(y^+_\times) = 10.48 - 42.0\,\Reytau^{-1/4}\quad ,
\end{equation}
with the corresponding $\langle uu\rangle^+_\times$ following from equation (\ref{uuininf}). To guide the eye, the points $(y^+_\times, \langle uu\rangle^+_\times)$ are marked by $\blacklozenge$'s for the profiles of figure \ref{fig3}.

The adoption of a constant $y^+_\times$ for all $\Reytau$ means, that the inner expansion, which cannot end at a finite value of the inner coordinate, has to include the logarithmic decay beyond the cross-over point $y^+_\times$. Hence, the complete inner expansion is obtained by adding a corner function (\ref{corner}) to $\langle uu\rangle^+_{\mathrm{wall}}$ of equation (\ref{uuwall}) :
\begin{equation}
\label{uuin}
\langle uu\rangle^+_{\mathrm{in}} = \langle uu\rangle^+_{\mathrm{wall}} + \mathcal{C}\big(y^+; y^+_\times, \Delta S, 2\big)
\end{equation}
and its limit for $y^+ \gg y^+_\times \gg 1$ is
\begin{align}
\langle uu\rangle^+_{\mathrm{in}}\,(y^+ \gg y^+_\times \gg 1) =& \,S_{\mathrm{in}}\,\big[\ln y^+ - \ln y^+_\times \big] + \langle uu\rangle^+_\times \,+\, \mathrm{H.O.T.} \nonumber \\ \mathrm{with}\quad
S_{\mathrm{in}} =& \,S_{\mathrm{wall}} + \Delta S
\label{uuininf}
\end{align}
where the logarithmic slope $S_{\mathrm{in}}$ of the asymptotic inner loglaw (\ref{uuininf}) must be determined by matching to the outer expansion in section \ref{sec3}. At this point it can only be said that it must
be negative to form an outer peak at large $\Reytau$. Furthermore, it \textbf{must go to zero} at infinite $\Reytau$, as shown in figure \ref{fig3}, because $\langle uu\rangle^+$ near the wall has been shown to remain finite for all $\Reytau$. This implies that $\Delta S$ is of the form $\Delta S = -0.85 + \Delta S'$, with $\Delta S'$ vanishing for $\Reytau\to\infty$,  to compensate the $\mathcal{O}(1)$ contribution to $S_{\mathrm{wall}}$ in equation (\ref{uuwallcof}).

Finally, as all the terms of equation (\ref{uuininf}) must be matched to the small-$Y$ limit of the outer expansion, established in the next section \ref{sec3}, the limit (\ref{uuininf}) of $\langle uu\rangle^+_{\mathrm{in}}$ must also be the common part of inner and outer expansions
\begin{equation}
\label{uucp}
\langle uu\rangle^+_{\mathrm{cp}} = \langle uu\rangle^+_{\mathrm{in}}\,(y^+ \gg y^+_\times \gg 1) \quad .
\end{equation}

\section{\label{sec3}The outer and composite expansions of $\langle uu\rangle^+$}

Moving on to the outer expansion, it is written as a logarithmic part matching the common part (\ref{uucp}) for small $Y$, plus a wake part $\mathcal{W}(Y)$, which goes to zero for $Y\to 0$ and satisfies the symmetry condition on the centerline $Y=1$
\begin{equation}
\label{uuout}
\langle uu\rangle^+_{\mathrm{out}} = S_{\mathrm{out}}\,\ln\left[\Reytau Y \left(1 - \frac{Y}{2}\right)\right] -
S_{\mathrm{out}}\,\ln y^+_\times + \langle uu\rangle^+_\times + \mathcal{W}(Y) + \mathrm{H.O.T.}
\end{equation}
The matching of outer and inner expansions furthermore requires $S_{\mathrm{out}}$ to be identical to $S_{\mathrm{in}}$ in equations (\ref{uuininf}, \ref{uucp}).

To obtain the logarithmic slope $S_{\mathrm{out}} = S_{\mathrm{in}}$, the outer expansion (\ref{uuout}) is evaluated at $Y=1$ and identified with the fit (\ref{CLfit}) of centerline data
\begin{align}
\label{uuCL}
\langle uu\rangle^+_{\mathrm{CL}} =& \,S_{\mathrm{out}}\,\ln\left(\frac{\Reytau}2\right) -
S_{\mathrm{out}}\,\ln y^+_\times + \langle uu\rangle^+_\times + \mathcal{W}(1) \\
=&\, 0.55 + \big[0.1007 + 33\,\Reytau^{-1/4}\big]^{-1} \quad .
\label{CLfit}
\end{align}
As seen in figure \ref{fig3}, this fit reproduces the channel and Superpipe centerline data up to $\Reytau = 10^5$, and increases to $\langle uu\rangle^+_{\times} = 10.48$ at infinite $\Reytau$, such that $\langle uu\rangle^+_{\mathrm{out}}$ becomes a simple constant throughout the channel or pipe. Note that the fit (\ref{CLfit}), which relies strongly on the Superpipe data of \citet{Hultetal12} (see fig. \ref{fig3}), implies that differences are small between the outer expansions for pipe and channel (see comments in section \ref{sec4}).

What is still missing for the determination of the outer logarithmic slope $S_{\mathrm{out}}$ is the wake function $\mathcal{W}(Y)$. The fit with the requisite properties
\begin{equation}
\label{Wfit}
\mathcal{W}(Y) = S_{\mathrm{out}}\,\ln\left[\frac{1}4 + \frac{3}4\,(1-Y)^2\right]
\end{equation}
has again been developed from both channel DNS and Superpipe NSTAP data. The resulting outer logarithmic slope is found to scale as
\begin{equation}
\label{Soutfit}
S_{\mathrm{out}} = \sigma \ln^2\Reytau \,\Reytau^{-1/4}\quad ,
\end{equation}
with $\sigma$ a weak function of $\Reytau$, varying between -0.19 and -0.15 in the interval $\Reytau \in [10^3, 10^{10}]$. While the $\Reytau^{-1/4}$ dependence is directly related to the scaling of $\langle uu\rangle^+_{\mathrm{in}}$, obtained without model assumptions, the factor $\ln^2\Reytau$ in equation (\ref{Soutfit}) may depend on the details of how $S_{\mathrm{out}}$ has been determined. However, as long as $\langle uu\rangle^+_\times$ remains finite, $S_{\mathrm{out}}$ must go to zero for $\Reytau \to \infty$.\newline

\begin{figure}
\center
\includegraphics[width=0.6\textwidth]{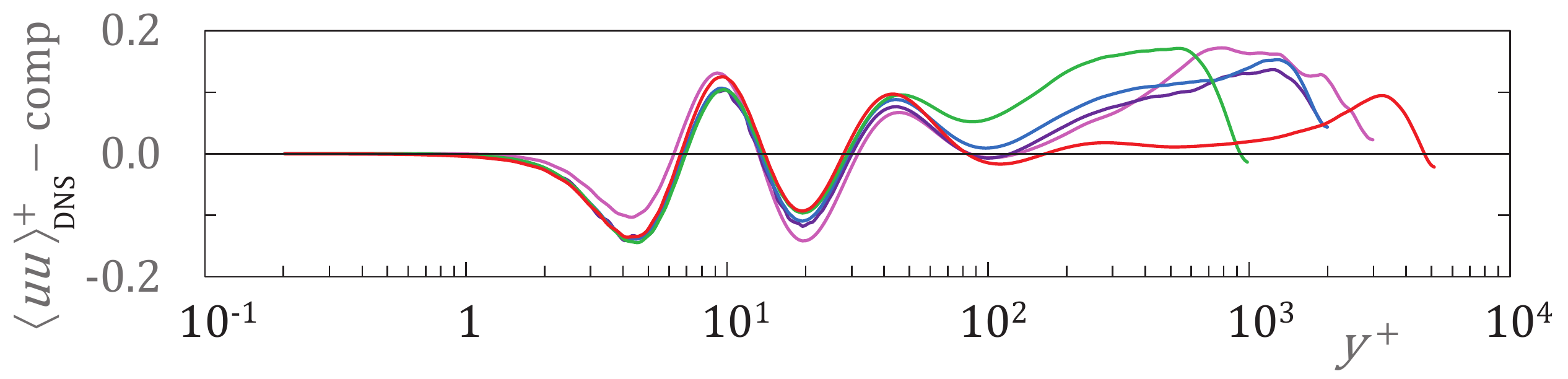}
\caption{\label{fig4} Difference between the DNS profiles $\langle uu\rangle^+$ of table \ref{TableDNS} and their complete composite fit (\ref{uucomp}).}
\end{figure}

With the determination of $S_{\mathrm{out}}$, the composite expansion
\begin{equation}
\label{uucomp}
\langle uu\rangle^+_{\mathrm{comp}} = \langle uu\rangle^+_{\mathrm{in}} + \langle uu\rangle^+_{\mathrm{out}} - \langle uu\rangle^+_{\mathrm{cp}}
\end{equation}
is complete up to $\mathcal{O}(\Reytau^{-1/4})$, and the different parts are given by equations (\ref{uuin}), (\ref{yx}), (\ref{uucp}), (\ref{uuout}) and (\ref{Wfit}).
This final result is first compared in figure \ref{fig4} to the DNS data of table \ref{TableDNS}, and the differences between composite expansion and DNS are seen to be at most $1.5\%$ of the inner peak height (\ref{uuIP}). It is noted, that in the region $0 \leq y^+ \lessapprox 10^2$ the difference between DNS and composite expansion is due to an imperfect ``hump'' function (\ref{Hump}). However, no improvement is pursued here, as the deviations from DNS are barely larger than the line thickness in figure \ref{fig1} and no additional insight would be gained from a more complex $\mathcal{H}$.

The complete composite expansion (\ref{uucomp}) allows the extrapolation of $\langle uu\rangle^+$ to truly large $\Reytau$, as shown in figure \ref{fig3}. The first thing to notice in this figure, is the close correspondence between composite profiles and both DNS and experimental profiles over the entire $\Reytau$ range of $10^3$ to $10^5$. Only the Superpipe data of \citet{Hultetal12}, deviate progressively from the composite expansion below $Y\approx 0.006$, corresponding to a wall distance of about $0.4$mm (see also fig. \ref{fig2}), but to this author's knowledge, no corrections have been devised for these NSTAP data.

The figure also shows the ``skeleton'' of the composite expansion: the fan of logarithmic asymptotes (\ref{uuwallinf}) of $\langle uu\rangle^+_{\mathrm{wall}}$ (\ref{uuwall}) and the corresponding asymptotic overlap loglaws (\ref{uuininf}), together with their intersections (\ref{yx}), marked by $\blacklozenge$'s. Note already the considerable difference, at the lower Reynolds numbers, between the composite expansion and its asymptotic logarithmic ``skeleton'', to be commented in the following section \ref{sec4}.

\section{\label{sec4}Discussion and outlook}

The principal features of the $\langle uu\rangle^+$ profiles shown in figure \ref{fig3} lead to the following comments and conclusions :
\begin{enumerate}
\item It has been demonstrated, that an unlimited growth of the inner $\langle uu\rangle^+$ peak with $\Reytau$ is incompatible with the channel DNS profiles in table \ref{TableDNS}. The model-free analysis of these profiles in section \ref{sec2} shows that the inner peak $\langle uu\rangle^+_{\mathrm{IP}}$ remains finite in the limit of infinite $\Reytau$ and decreases from there as $\Reytau^{-1/4}$ or $1/\ln\Reytau$. Based on the theoretical arguments of \citet{chen_sreeni2021}, only the $\Reytau^{-1/4}$ scaling (equ. \ref{uuIP}) has been pursued here. In other words, the unlimited growth of $\langle uu\rangle^+_{\mathrm{IP}}$ predicted by the attached eddy model \citep{MarusicMonty19} is not borne out by the data, even though its prediction falls within the data uncertainty in the Reynolds number range of figure \ref{fig2}.
\item The possibility of an unlimited growth of the \textbf{outer} peak with $\Reytau$ appears very unlikely, considering that the cross-over location $y^+_\times$ from the wall loglaw (\ref{uuwallinf}) to the overlap loglaw not only scales on inner units, but remains constant over the $\Reytau$ range where laboratory data are available. On figure \ref{fig3}, an unlimited growth of the outer peak would require the vertical line connecting the intersections $\blacklozenge$ of wall and overlap loglaws to sharply bend to the right above the intersection for the highest NSTAP Reynolds number.
\item Support for the present composite profiles of figure \ref{fig3} also comes from the height of all distinct outer peaks reported in the literature, which are systematically below the corresponding intersections $\langle uu\rangle^+_\times(\Reytau)$ (equ. \ref{yx}). The same is true for the outer peak correlations $2.82+0.42\,\ln\Reytau$ and $0.33+0.63\,\ln\Reytau$ of \citet{Pullin2013} up to $\Reytau$'s well above $10^6$.
\item With both inner and outer peaks of $\langle uu\rangle^+$ finite, a simple geometric argument, already brought up by \citet{MonkNagib2015}, rules out a Reynolds-independent slope of the overlap loglaw. The attached eddy model may nevertheless be useful for intermediate Reynolds numbers, as the universal logarithmic slope of -1.26 inferred from the model and indicated in figure \ref{fig3} by the dash-dotted line, is relatively close to the slopes of the overlap loglaws below $\Reytau$ of $10^5$, which vary between -1.6 and -1.17 .
\item The switch-over from the wall loglaw to the overlap loglaw at a \textbf{fixed} value of $y^+_\times \approxeq 470$, identified in the present paper, is reminiscent of the change of logarithmic slope from $1/0.398$ to $1/0.42$ at $y^+ = 624$, found by \citet{Monk21} in the mean velocity profile. A connection between the two observations is likely, with the chain of cause and effect going from the normal stresses, $\langle uu\rangle^+$ in particular, via the Reynolds stress $\langle uv\rangle^+$  to the mean velocity.
\item The transition from the wall loglaw to the overlap loglaw may also be related to the topological change of the velocity-vorticity correlation found by \citet{chen_hussain_she2019} to occur around $y^+ \approx 110$, but the possibility of an opposite wall effect, discussed by \citet{Monk21}, remains open.
\item Also open is the question whether the $\langle uu\rangle^+$ composite expansion developed here is universal or not. For the near-wall region in figure \ref{fig1}, the present composite expansion relies entirely on channel DNS, while at the higher $\Reytau$, the outer part of the Superpipe profiles of \citet{Hultetal12} has helped guide the expansion. The close correspondence in figure \ref{fig3} between the channel DNS and Superpipe profiles for $\Reytau$ of 1985 and 5411 suggests that the differences between channel and pipe are small. It would however be surprising, if there were no differences at all, at least in the outer $\langle uu\rangle^+$ expansion, as there are strong indications \citep{Monk21} that the outer mean velocity expansions, in particular the K\'arm\'an ``constants'', are different for channel and pipe.
\item Finally, it must be reiterated, that determining the slopes of loglaws, which are inherently asymptotic laws, by fitting tangents to finite Reynolds number data is hazardous. As illustrated in figure \ref{fig3}, only at the highest NSTAP $\Reytau$ of around $10^5$ does the overlap loglaw start to go through the data ! This is the same conclusion as the one reached by \citet{Monk21}, who found that the mean velocity indicator function $y^+(\mathrm{d}U^+/\mathrm{d}y^+)$ starts to reach the correct plateau only beyond a $\Reytau$ of around $10^5$. Up to such high $\Reytau$, the development of proper asymptotic expansions is indispensable.
\end{enumerate}

\begin{acknowledgments}
I am grateful to Katepalli ``Sreeni'' Sreenivasan, Xi Chen and Hassan Nagib for their helpful comments and encouragement.
\end{acknowledgments}

Declaration of Interests. The author reports no conflict of interest.
\appendix
\section{\label{App}The ``Musker-like'' fit for the near-wall $\langle uu\rangle^+$ profile and other fits}

Following the idea of \citet{Musker79}, the $\mathcal{O}(1)$ part of the inner (near-wall) $\langle uu\rangle^+$ profile is approximated analytically by the integral of $\dd \mathcal{M}_2/\dd y^+ = 2 p_0\, y^+\,[p_1 + p_2(y^+)^2][p_1 + p_3(y^+)^2 + (y^+)^4]^{-1}$, where the subscript ``2'' indicates that the leading term of the Taylor expansion of $\mathcal{M}_2$ around $y^+=0$ is $\propto(y^+)^2$. The result of the integration is
\begin{align}
\label{M2}
&\mathcal{M}_2\left(y^+; p_0, p_1, p_2, p_3\right) =
\frac{p_0 (2p_1-p_2p_3)}{\sqrt{4p_1-p_3^2}}\,\left\{\arctan \frac{p_3+2(y^+)^2}{\sqrt{4p_1-p_3^2}} -
\arctan \frac{p_3}{\sqrt{4p_1-p_3^2}}\right\}
\nonumber \\
&+\frac{p_0 p_2}{2}\,\ln\left[1 + \frac{p_3 (y^+)^2}{p_1} + \frac{(y^+)^4}{p_1}\right] \quad ,
\end{align}
where the parameters $p_0$ ... $p_3$ are determined by the boundary conditions. For large $y^+$, $\mathcal{M}_2$ asymptotes to the loglaw
\begin{equation}
\label{M2log}
\mathcal{M}_2(y^+\gg 1) = \frac{p_0 p_2}{2}\,\Big\{4\,\ln(y^+)-\ln(p_1)\Big\} + \frac{p_0}{\sqrt{4p_1-p_3^2}}\,\left\{\frac{\pi}{2} - \arctan \frac{p_3}{\sqrt{4p_1-p_3^2}}\right\}
\end{equation}
and near the wall it has the Taylor expansion
\begin{equation}
\label{M2Taylor}
\mathcal{M}_2(y^+\to 0) = p_0 (y^+)^2 + \frac{p_0 (p_2 - 2p_3)}{2 p_1}\,(y^+)^4 + ...
\end{equation}

Like the original mean velocity Musker profile, the profile (\ref{M2})
misses a ``hump'' centered around $y^+ =\mathcal{O}(10)$. As in \citet{Monk21}, it is modelled by the ``hump'' function
of \citet{NagibChauhan2008}
\begin{equation}
\label{Hump}
\mathcal{H}(y^+; h_1, h_2, h_3) = h_1\,\exp\left[- h_2\,\ln^2(y^+/h_3)\right]\quad .
\end{equation}

Finally, the smooth transition between two logarithmic laws with different slopes in a variable $\eta$ is fitted by the corner function
\begin{equation}
\label{corner}
\mathcal{C}(\eta; \eta_c, c, m) = \frac{c}{m}\,\ln\left[1 + \left(\frac{\eta}{\eta_c}\right)^m\right]\, \rightarrow \,\left\{
\begin{array}{ll} 0 & \textrm{for}\quad \eta\ll\eta_c \\ c\,[\ln\eta-\ln\eta_c] & \textrm{for}\quad \eta\gg\eta_c \end{array} \right.
\end{equation}
where the sharpness of the corner at $\eta_c$ is governed by the parameter $m$.

\bibliographystyle{jfm}
\bibliography{Turbulence}

\end{document}